\documentclass[12pt,preprint]{aastex}
\usepackage{graphicx}

\shorttitle{Granular-scale Flux Emergence and Cancellation}
\shortauthors{Lim et al.}

\begin{document}

\title{Photospheric Signatures of Granular-scale Flux Emergence and Cancellation at the Penumbral Boundary}

\author{Eun-Kyung Lim, Vasyl Yurchyshyn, Valentyna Abramenko, Kwangsu Ahn, Wenda Cao, and Philip Goode}
\affil{Big Bear Solar Observatory, New Jersey
Institute of Technology, 40386 North Shore Lane, Big Bear City, CA
92314-9672, USA}
\email{eklim@bbso.njit.edu}

\begin{abstract}
We studied flux emergence events of sub-granular scale in a solar active region. New Solar Telescope (NST) of Big Bear Solar Observatory made it possible to clearly observe the photospheric signature of flux emergence with very high spatial (0\arcsec.11 at 7057~\AA) and temporal (15~s) resolution. From TiO observations with the pixel scale of 0\arcsec.0375, we found several elongated granule-like features (GLFs) stretching from the penumbral filaments of a sunspot at a relatively high speed of over 4~km~s$^{-1}$. After a slender arched darkening appeared at a tip of a penumbral filament, a bright point (BP) developed and quickly moved away from the filament forming and stretching a GLF. The size of a GLF was approximately 0\arcsec.5\, wide and 3\arcsec\, long. The moving BP encountered nearby structures after several minutes of stretching, and a well-defined elongated shape of a GLF faded away. Magnetograms from SDO/HMI and NST/IRIM revealed that those GLFs are photospheric indicators of small-scale flux emergence, and their disappearance is related to magnetic cancellation. From two well-observed events, we describe detailed development of the sub-structures of GLFs, and different cancellation processes that each of the two GLFs underwent.
\end{abstract}


\section{Introduction}
Flux emergence is the main source for the formation of magnetic structures in the photosphere. Magnetic flux tube emerges from the convective zone through the photosphere by means of magnetic buoyancy due to the lower gas pressure inside the flux tube \citep{Par55}. Instead of an abrupt emergence of a large bundle of flux tube, \citet{Zwa87} suggested that a large-scale magnetic structure forms through the coalescence of numerous small-scale magnetic elements. \citet{Str99} observed that small-scale magnetic flux subsequently emerges to form an active region, and 3D magnetohydrodynamics simulations by \citet{Che08,Che10} showed that small-scale magnetic elements uplifted by buoyant force gradually coalesce into larger structures such as pores and sunspots. Recently, \citet{Sch10,Sch10b} observed the formation of an elongated granule at the vicinity of a sunspot with both photometric and spectropolarimetric data and suggested that emerging loops on small scales may play an important role in a developing sunspot with a penumbra.

Indeed, to study small-scale magnetic elements emerging in an active region is important to understanding the ovarall magnetic structures of a sunspot. Especially, flux emergence in a moat region has become an important topic in the context of the origin of fine-structures such as penumbral filaments and moving magnetic features (MMFs). Such fine-structures are believed to be magnetically connected with the large-scale active region field in the form of a sea-serpent configuration \citep[etc.]{Har73, Sch02, Wei04}, and hence observational properties of those structures, such as magnetic configuration, magnetic flux content, velocity, and the direction of movements \citep{Lee92,Shi01,Yur01,Zha03,Hag05} have become critical constraints of sunspot models.

Although we have acquired substantial knowledge about small-scale emerging flux owing to recent improvements in polarimetric sensitivity \citep{Lit96, Dep02, Lit08}, it seems that many small-scale magnetic structures are still not fully detected and/or resolved \citep[see for example,][]{ste11}. On the other hand, the most up-to-date photospheric photometric data show fine-structures on sub-granular scale that seem to be associated with magnetic fields \citep{Goo10b}. Studying such photospheric fine-scale structures could provide a tool for understanding the behavior of unresolved magnetic fields. Emerging flux will is associated with a photospheric counterpart, such as elongated granules when a loop top passes the photosphere \citep{Bra64, Sch10}, and many studies investigated morphological properties of such photospheric features in detail. For instance, \citet{Str99} studied the pattern of granular motion in an emerging active region utilizing both continuum intensity and circular polarization data, and noted several interesting phenomena, such as a transient line-center (or continuum) darkening associated with upflows later followed by bright grains. Although the emergence events they studied were relatively large ($\geq1$~Mm), as compared to those addressed in recent spectropolarimetric works, this study provided insights on photospheric signatures associated with the emergence of the magnetic field. Additional attempts utilizing up-to-date data showing features on sub-granular scale, even without detecting magnetic field, are worth performing to understand the process of flux emergence on small spatial scales.

Studies focused on photospheric fine-scale counterparts of the emerging flux in the moat region are not numerous, so far. Better understanding of fine-scale features in the photosphere produced by emerging flux will also provide important information on the magnetic structure itself. Our aim is to fill this gap by studying photospheric sub-granular scale features that are believed to be signatures of flux emergence. Here we present recent observations of photospheric granular motion in an active region associated with small-scale emerging flux observed with both high spatial resolution and time cadence. The data have been acquired with a 1.6~m clear aperture New Solar Telescope \citep[NST;][]{cao10a,Goo10} operating at the Big Bear Solar Observatory (BBSO).

\section{Observations}
\begin{figure}[tbp]
\begin{center}
    \epsscale{0.8}
    \plotone{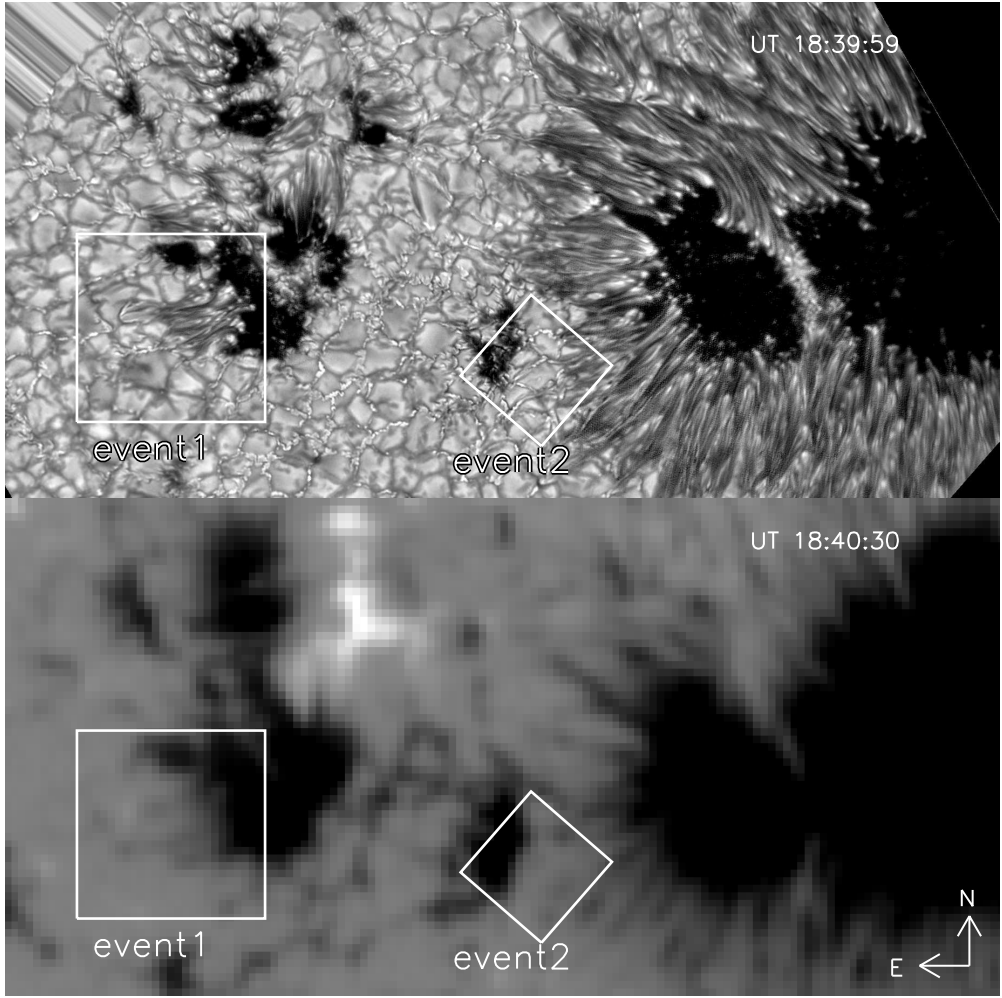}
    \caption{Images of studied active region. Top: TiO image taken on September 17, 2010. Bottom: HMI line-of-sight magnetogram taken at the nearest time to TiO data (top). Both TiO and HMI images are spatially co-aligned, and locations of the two events of our interest are presented as white squares. The field of view of each panel is $63\arcsec\, \times31\arcsec\,.5$}\label{overall}
\end{center}
\end{figure}

On September 27, 2010, the NST observed a small fraction of active region NOAA 11109 using NST Filter Imagers \citep{cao10b} including a broadband TiO filter with a center wavelength of 7057~\AA\, and a 0.25~\AA\, bandpass Zeiss Lyot birefringent H$\alpha$ narrow band filter. Both TiO and H$\alpha$ data cover a 2.5 hour period from 17:22:54~UT to 19:51:39~UT. During this time, active region NOAA 11109 was located near the central meridian, 20 degrees north of the equator. The data were taken with an aid of adaptive optics \citep{cao10a} in many bursts of TiO and H$\alpha$ images. Each TiO burst consisting of 70 images, was taken within a 15~s interval with an exposure time of 1~ms. We applied the Kiepenheuer-Institut Speckle Interferometry Package \citep[KISIP;][]{Wog08} and obtained speckle reconstructed images with a field of view (FOV) of $69\arcsec.8 \times 69\arcsec.8$, and a pixel size of 0\arcsec.0375.

H$\alpha$ observations were performed by sequentially shifting the central wavelength of the filter to $-0.75$~\AA, H$\alpha$ line center, and $+0.75$~\AA. Different exposure times were employed based on the central wavelength of the filter, and the time cadence of bursts was set to be around 10~s. All H$\alpha$ bursts were also speckle reconstructed using the KISIP code, and the reconstructed images have a FOV of $76\arcsec.4 \times 76\arcsec.4$ with a pixel scale of $0\arcsec.075$. Both TiO and H$\alpha$ images were carefully aligned and de-stretched to remove resional image jitter and distortions.

To analyze magnetic field distribution, we used full-disk magnetograms recorded with a 45~s cadence by the Helioseismic and Magnetic Imager (HMI) on board the \emph{Solar Dynamics Observatory} (\emph{SDO}) data and NST's InfraRed Imaging Magnetograph \citep[IRIM;][]{cao11} data. The spatial resolution of {HMI} is 1\arcsec\,. Most of observed events show a very fine structuring of less than 1\arcsec\, in width. In order to detect magnetic fields of these very small structures seen in TiO observations, we used line-of-sight magnetograms taken by IRIM as complementary data. IRIM is an imaging solar spectro-polarimeter that uses a pair of Zeeman sensitive Fe~{\footnotesize I} lines present in the near infrared at 15648.5~\AA\, and 15652.9~nm. The IRIM optical layout included a 25~\AA\, interference pre-filter, a 2.5~\AA\, birefringent Lyot filter and a 0.1~\AA\, Fabry-P\'{e}rot etalon. With the aid of adaptive optics, we obtained circular polarization IRIM data with spatial and time resolution of 0\arcsec.2\, and 43~s, respectively. Since the spatial and temporal coverage of IRIM data was lower than that of the TiO and H$\alpha$ data, the IRIM data was used as complementary to the {SDO}/HMI. Because of unfinished polarization calibration, there was a loss of circular polarization (Stokes-V) signal due to non-negligible cross-talk in IRIM. Nevertheless, the comparison between IRIM data and SDO/HMI showed a consistent distribution of the line-of-sight magnetic field, and IRIM magnetograms are reliable for the morphological study.

From the 2.5-hour TiO dataset, we selected a subregion east of the main negative polarity sunspot. Figure~\ref{overall} shows the region of our interest. This area contained a group of small pores and a part of the penumbra of the main sunspot. The magnetic field of pores including the main spot have negative polarity except for one small positive-polarity pore. Most of granule-like features (GLFs) developed from the boundary of the penumbral region of the main spot, and some of them from the easternmost group of negative pores in Figure~\ref{overall}. We selected two GLF events that were well-observed and showed interesting, associated chromospheric dynamics (event1 and event2 denoted in Figure~\ref{overall}). With those events, we analyzed the morphological evolution during their lifetime along with the associated transverse plasma flows obtained from the time-distance (x-t) plot from TiO data. The corresponding magnetic features and associated chromospheric dynamics were examined from the SDO/HMI, BBSO/IRIM and BBSO/H$\alpha$ data, respectively.

\section{Results}
Two events of formation of elongated GLFs were well-recorded in TiO data and {SDO}/HMI and BBSO/IRIM magnetograms. Each of these events was associated with different types of chromospheric dynamics as observed in H$\alpha$ data. Below, we present a detailed description of evolution of their morphologies, magnetic structures and associated chromospheric features.

\subsection{Event 1}\label{sec.ev1}

\begin{figure}[tbp]
\begin{center}
    \epsscale{1.}
    \plotone{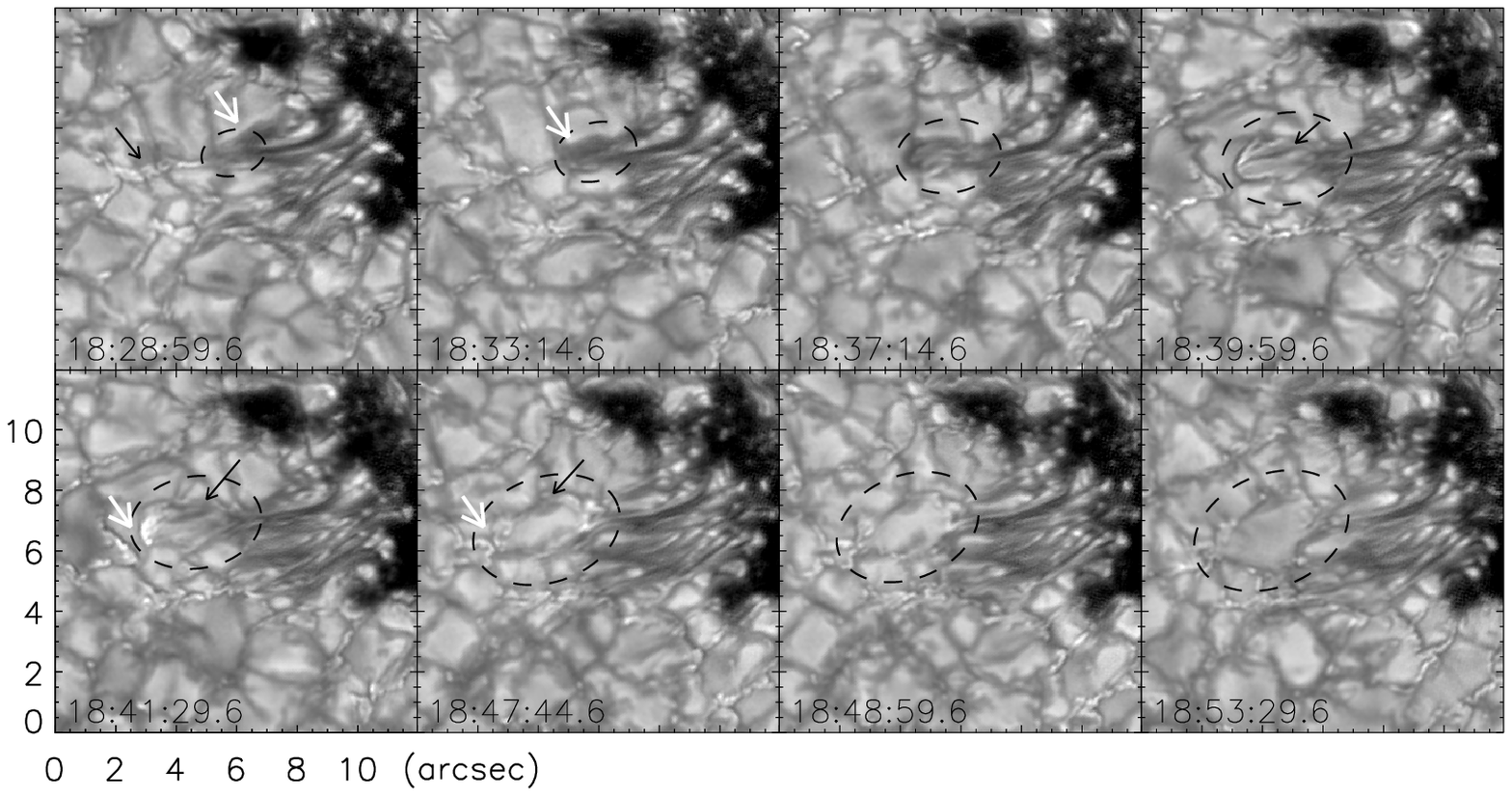}
    \epsscale{1}
    \plotone{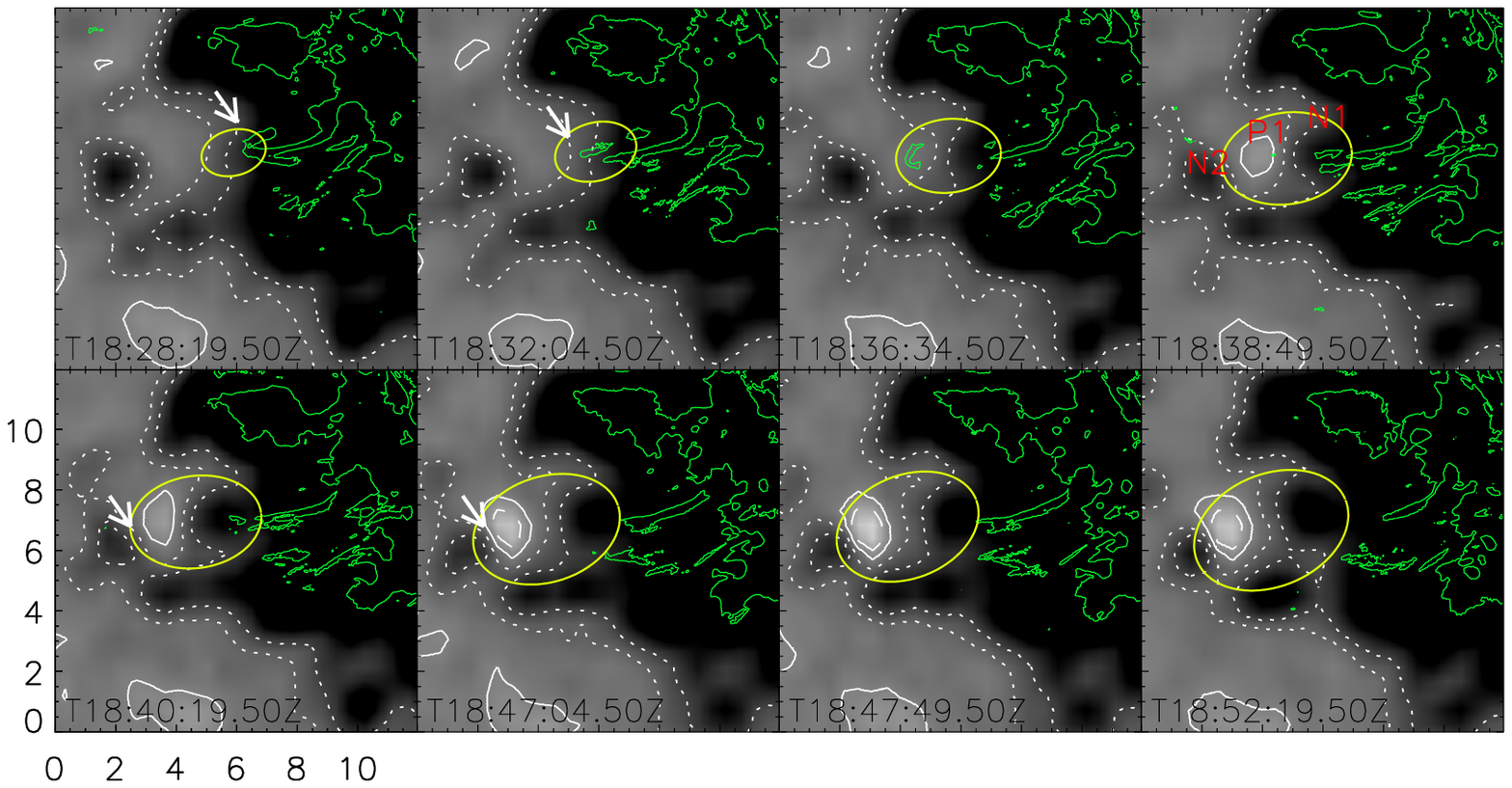}
    \caption{Top: time series of TiO observations taken on 27 Sept. 2010. Black dashed ovals indicate the GLF of our interest, white arrows the farther footpoint of the GLF from the pore, and black arrows the thread structure of the GLF. [\emph{Mpeg file is available in the electronic edition of the Journal.}] Bottom: line-of-sight {SDO}/HMI magnetograms (background) spatially and temporally co-aligned with TiO data (green contours). White contours indicate magnetic flux, dotted curves corresponding to the magnetic flux of (-50, -100 G), solid curve to (0 G), and the dashed curve to (+50 G). Same white arrows and ovals shown in the top panel are also marked in the magnetograms. P1 and N1 denote positive and negative magnetic flux at footpoints of the GLF, and N2 denotes a preexisting negative flux.}\label{mmf1}
\end{center}
\end{figure}

This event occurred near the negative-polarity pore shown in Figure~\ref{mmf1}, which is a member of the negative pore group. This pore had an elongated shape with its longer side measuring 9\arcsec\,. Its left side is far from the sunspot and a proto-penumbra was developing at this location. The location of an elongated GLF is denoted by dashed ovals in all panels of Figure~\ref{mmf1}. An animation file can be found in the electronic edition of this paper, and on the BBSO Web site\footnote{http:\///bbso.njit.edu\//nst\_\,gallery.html}. The GLF first appeared at 18:28:59 UT as a faint darkening at the tip of a penumbral filament (white arrow), and then moved away from the penumbra acquiring a bean-like shape by 18:37:14 UT. By 18:39:59 UT, the right tip of the GLF was still connected to the penumbral filament, while the left, moving footpoint had a sharp, dark boundary. In addition to the sharp, dark boundary, the figure reveals the existence of dark and bright threads inside the GLF aligned with the direction of its elongation (black arrow at 18:39:59 UT). These threads were visible for about 5~minutes, then they merged into an intergranular lane as the GLF expanded (black arrows in the second row). It is note worthy that the GLF, which is fine-scale itself, consists of fine threads similar to a penumbral filament. Further discussions of this are in Section~4.

Note that there is a faint bright thread structure at the left side that is indicated by a black arrow in the first image, which was there before the GLF appeared. As the GLF developed, its moving footpoint approached and encountered the preexising bright thread becoming brighter (18:41:29 UT). After the encounter, these two structures faded and a normal granular cell pattern appeared instead. The slight brightening on the right tip of the GLF at 18:47:44~UT indicates the probable footpoint of this feature, and gives the impression that this GLF may be associated with a magnetic bipole. The brightness of the left footpoint of the GLF was greatest just before it encountered the preexisting thread, and no significant flare was observed during the encounter.

The bottom of Figure~\ref{mmf1} shows line-of-sight magnetograms taken by the {SDO}/HMI instrument and co-aligned with the TiO data (top). A comparison of TiO images and magnetograms shows that the formation of the GLF is related to the emerging and moving magnetic flux. At 18:28:19~UT, the darkening in the TiO image was co-spatial with the negative flux of the pore. Then as the GLF grew, the boundary of the negative flux protruded (N1) and the positive flux element (P1) appeared next to the protruding negative part (18:38:49~UT). The left footpoint of the GLF is co-spatial with the positive flux, while its right footpoint is associated with the negative protruding flux, and the longer axis of the GLF connected two magnetic elements (P1 and N1). Although the negative flux was not fully disconnected from the pore, the data support an emergence of a bipolar flux. As this presumed $\Omega$-loop emerged, the field lines connecting the opposing polarities passed through the photosphere producing dark and bright threads along the field apex, while the separation of opposite polarities results in the elongation of a granule.

\begin{figure}[tb]
\epsscale{1}
    \plotone{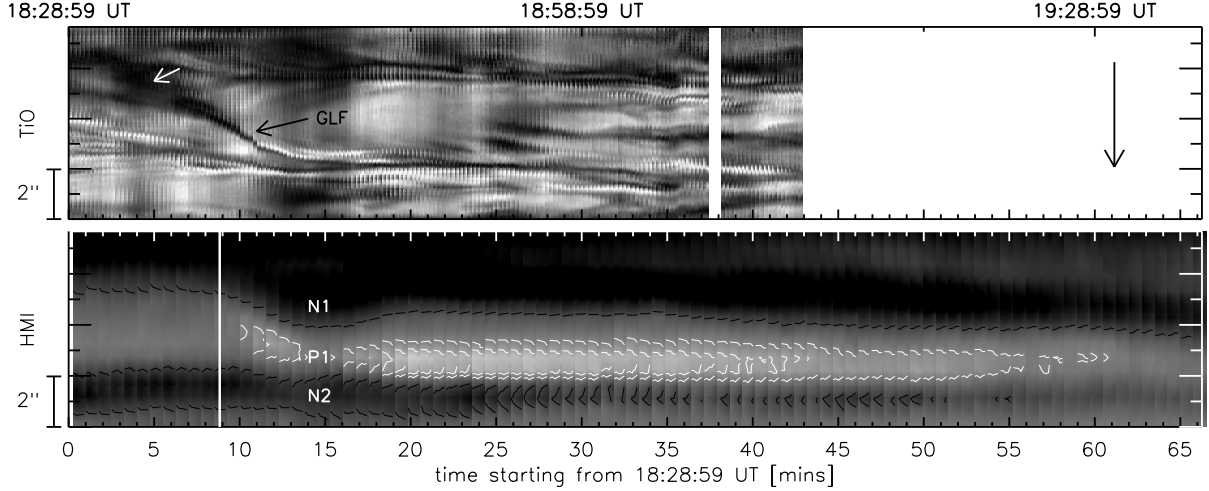}
    \caption{x-t plots obtained from TiO (top) and {SDO}/HMI (bottom) taken along the GLF's longer axis. Top : black curve pointed by black arrow (p1) is the path of the GLF's left footpoint, and the direction of its displacement is represented by a vertical arrow. Bottom : the vertical solid line indicates the time when the footpoint intensity begins to increase in both TiO and H$\alpha$ data (Figure~\ref{mmf1_flux}). Magnetic field strength is saturated at $\pm 200$~G, and contours indicate magnetic field strength of ($-100, 0, 50$~G). Note that the slit position was determined independently for two plots.}\label{mmf1_tslice}
\end{figure}

The time evolution of magnetic elements of the GLF during the entire period is well-displayed in x-t plots obtained from TiO and {SDO}/HMI data in Figure~\ref{mmf1_tslice}. The top panel shows the appearance of the darkening (white arrow), dark boundary of the moving footpoint (black arrow) followed by bright point (BP) at $(11,3)$ along the path of the left, moving footpoint of the GLF. Meanwhile, the bottom panel shows the emergence of both the positive flux element, P1, and the negative flux, N1, and the displacement of P1 outward from the pore. Most noticeable photospheric signatures of the GLF described above were observed during this period. The trace of the GLF's footpoint is seen as an inverse--S curve due to an acceleration between $t=8.5$ and $t=11$, and a deceleration after $t=11$. The average speed of the moving footpoint (P1) was estimated to be $5.0$~km~s$^{-1}$ during the acceleration period. Both magnetic configuration and apparent velocity consist with the type I MMFs \citep{Shi01}, and hence the result suggests that the observed GLF may be a photospheric counterpart of the MMF. Cancellation between P1 and preexisting negative flux, N2 is also shown. Although it is not clear when cancellation began, it seems that cancellation occurred for longer time period compared to the emergence. Recall that the cancellation appeared in TiO images as mixing of two bright structures (second row of the top of Figure~\ref{mmf1}), and even after the TiO structures disappeared, the magnetic elements were still seen in Figure~\ref{mmf1}.

\begin{figure}[tb]
\begin{center}
    \includegraphics[width=0.45\textwidth]{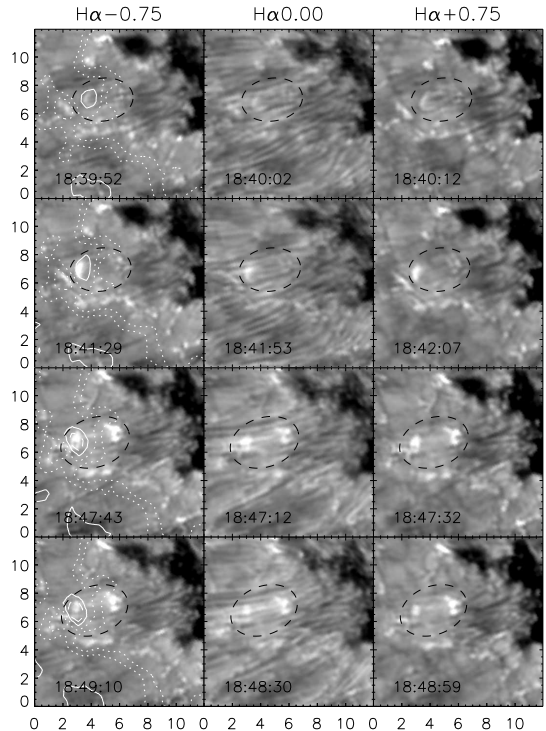}
    \caption{H${\alpha}$ observations taken at $-0.75, 0.00, +0.75$~\AA\, from line center at the same concurrently with TiO images.}\label{mmf1_hal}
\end{center}
\end{figure}

H$\alpha$ data showed no evidence for brightenings or jets associated with cancellation. Instead, faint brightenings at footpoints of the GLF were observed during the formation period of the GLF. Figure~\ref{mmf1_hal} shows H$\alpha$ blue-wing, center, and red-wing images taken at the nearest time to the TiO observations with the same FOV, as in Figure~\ref{mmf1}. Dark fibrils corresponding to the penumbral filaments were observed in all H$\alpha$ data at different wavelengths. A faint brightening appeared at the position corresponding to the left footpoint of the GLF at 18:39:52~UT, and another brightening at the right footpoint at 18:47:43~UT. These two brightenings are co-spatial with two footpoints of the GLF, and seem to develop due to emerging magnetic flux rather than cancellation.

\begin{figure}
\begin{center}
    \includegraphics[width=0.5\textwidth]{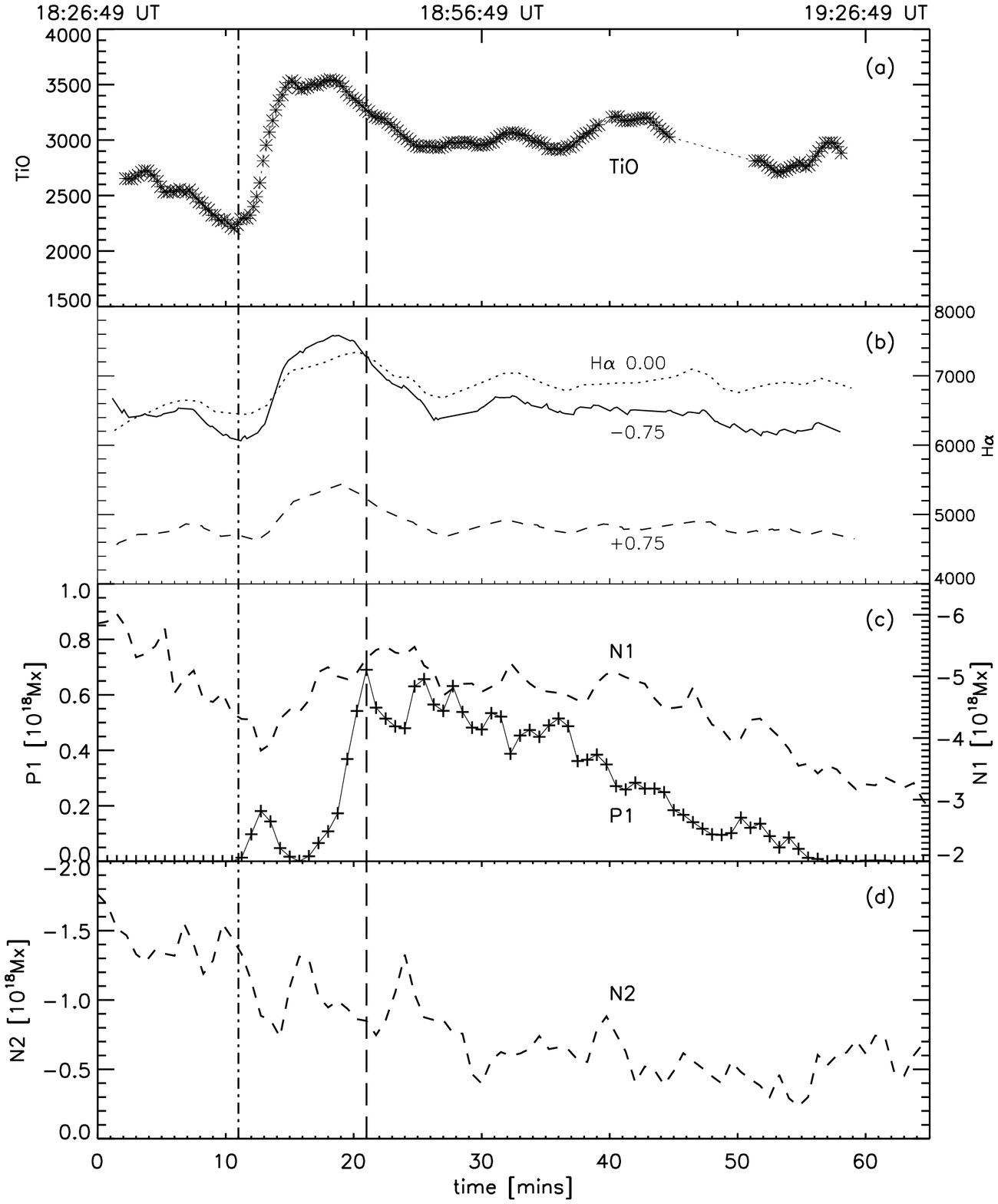}
    \caption{Time profiles of the mean intensity around the GLF's left footpoint obtained from TiO (a), and H$\alpha$ (b), time profiles of magnetic flux at P1 and N1 corresponding to the left and right footpoint of the GLF, respectively (c), and at N2 corresponding to the pre-existing negative flux (d).}\label{mmf1_flux}
\end{center}
\end{figure}

For further examination, the temporal change of the footpoint intensity in both TiO and H$\alpha$ data is compared with the flux variation at P1, N1, and N2 in Figure~\ref{mmf1_flux}. The mean intensity of TiO and H$\alpha$ data was obtained at the left footpoint of the GLF. It is clearly seen that the intensity profiles of TiO and H$\alpha$ (blue-wing) show almost the same pattern (a,b). Both intensity profiles begin to increase at $t=11$ (vertical dash-dot line), and this increase coinsides with the appearance of the positive flux P1 (c). The flux profile of P1 shows an increase between $t=11$ and $t=21$, then shows a decrease after $t=21$, with some fluctuations. Since the flux profile of N1 also increases during this period, this time period could be considered as the emerging phase of the bipolar flux. The steady decrease in flux profile of N2 (d) makes it difficult to determine the onset of cancellation. Instead, it seems that cancellation lasts for the entire period of the existence of P1. Note that all intensity profiles of TiO and H$\alpha$ increase during the emerging phase and begin to decrease even before P1 reaches its maximum peak. The plot shows that brightenings in TiO and H$\alpha$ are correlated with emergence of bipolar flux rather than cancellation.

\subsection{Event 2}
\begin{figure}[tb]
\begin{center}
    \includegraphics[height=0.5\textheight]{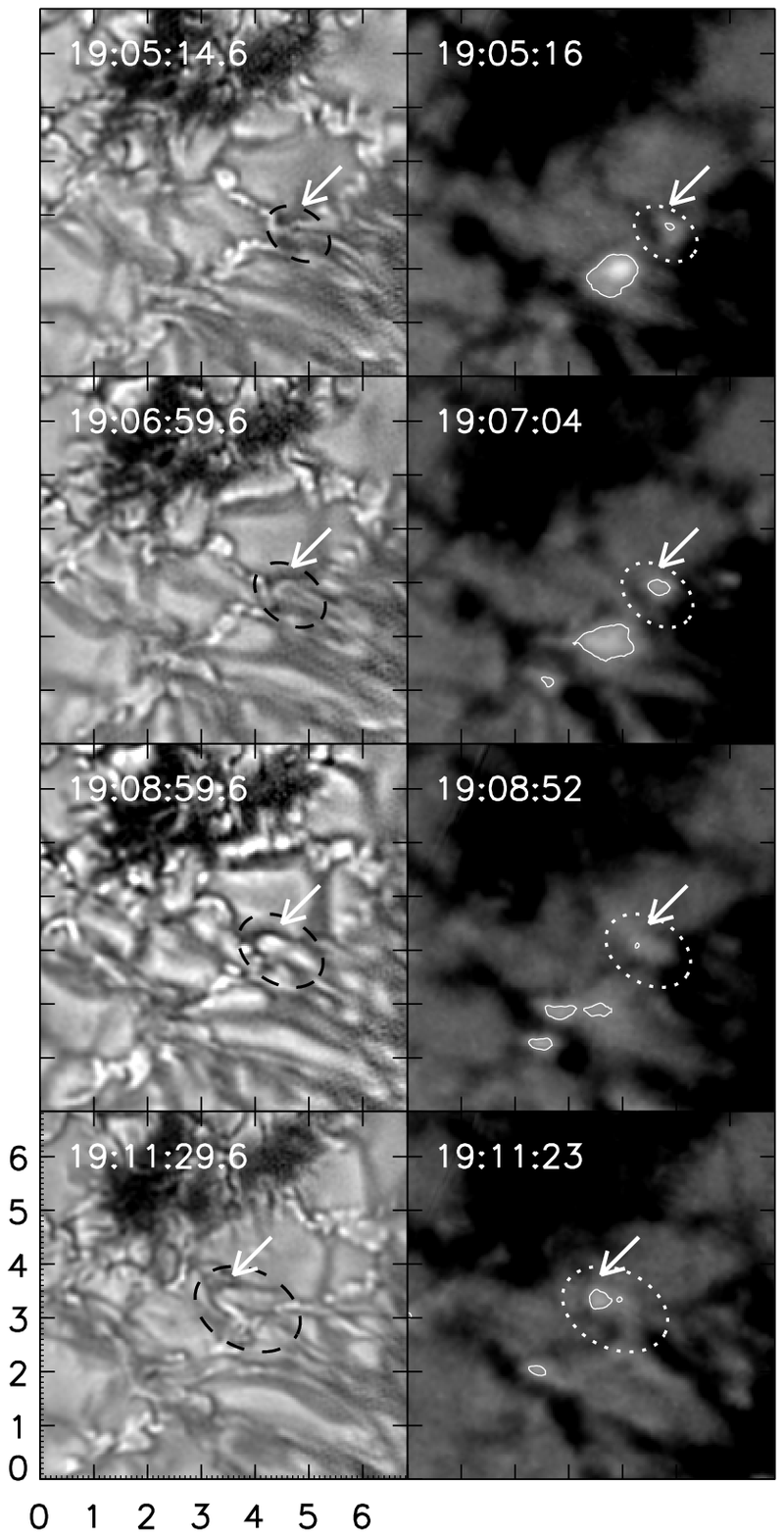}
    \includegraphics[height=0.5\textheight]{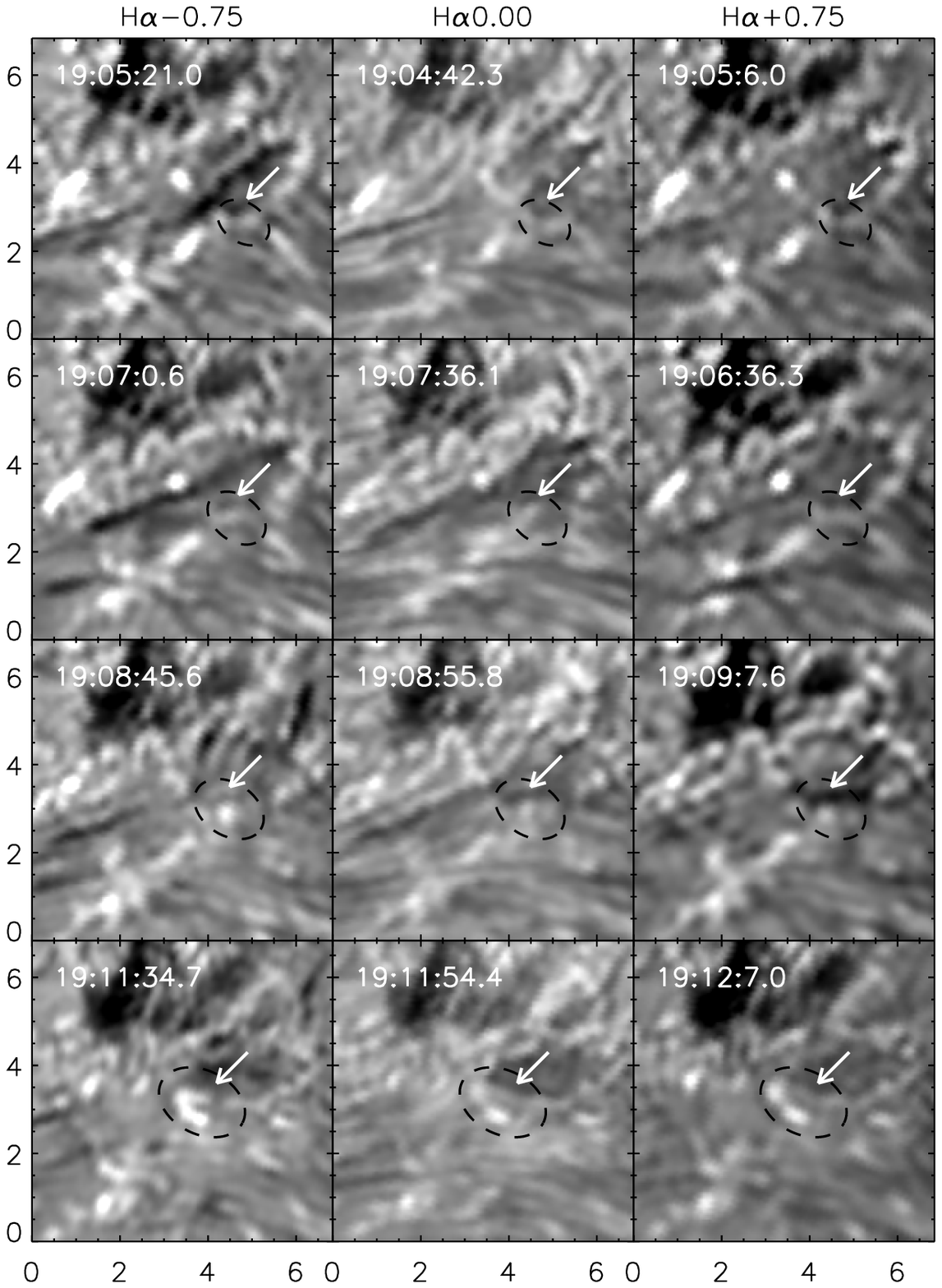}
    \caption{Left: TiO images showing the formation of the GLF (first column) and its associated Stokes-V maps obtained by the IRIM (second column). The white contour indicates a null in the Stokes V. Right: H$\alpha$ images taken in the blue-wing, center and red-wing of the line. All Stokes-V and H$\alpha$ images are carefully co-aligned with TiO data, and each arrow and oval indicates the position of the GLF.}\label{mmf2_tio.irim.hal}
\end{center}
\end{figure}

The second event also occurred at the penumbral filament of the main spot (Figure~\ref{overall}), and showed a similar behavior to that of the first event. TiO images in the left panel of Figure~\ref{mmf2_tio.irim.hal} show a part of the outer penumbra of the main spot in the lower-right corner and a small pore immediately adjacent to the penumbra in the upper-left corner. The pore was quite small, less than 5\arcsec\,, and only 2\arcsec\, away from the penumbra of the main spot. A penumbral filament indicated by an arrow in the 19:05:14 UT TiO image displays a well-defined structure consisting of a dark and bright threads. Interestingly, the co-aligned IRIM magnetogram show that the outermost tip of that structure is associated with the Stokes-V signal, but with the sign opposite to that of the main penumbral filament. Moreover, as the tip of this penumbral filament protruded toward the pore, the positive Stokes-V signal also followed the movement of the tip. Similar to the first event, an elongated GLF developed from the tip of the penumbral filament showing a sharp, dark boundary ahead of a BP (19:08:59~UT and 19:11:29~UT). Although no fine structure in the threads inside the GLF were detected in this event (probably due to its smaller size), the existence of a sharp, dark boundary at the moving front was clearly observed in both events. As was mentioned above, it seems that the formation of this GLF is associated with the appearance of a tiny positive flux adjacent to the negative flux of the sunspot penumbra. We were not able to detect any changes in the negative polarity penumbra that would indicate an appearance of a new magnetic flux element. This was probably because of its low magnetic strength, so that it remained masked by strong fields. A brightening at the GLF was observed from the H$\alpha$ data (right panel) which was particularly well-pronounced in the blue-wing images.

\begin{figure}[tb]
\begin{center}
    \includegraphics[width=0.9\textwidth]{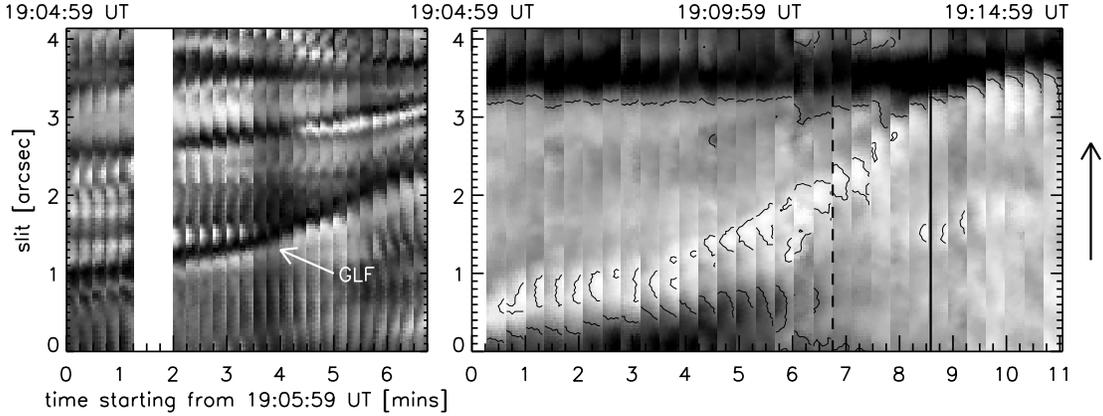}
    \caption{x-t plots obtained using TiO/IRIM data (left/right) measured along the path of the left footpoint of the GLF. The direction of the GLF's movement is denoted by a vertical back arrow. A white arrow (left panel) indicates the trace of the GLF, and black solid line (right panel) indicates the time at which the footpoint intensity in H$\alpha$ at all wavelengths (Figure~\ref{mmf2_ha.intflux}) begins to increase.
    }\label{mmf2_timeslice}
\end{center}
\end{figure}

\begin{figure}[tb]
\begin{center}
    \includegraphics[width=0.5\textwidth]{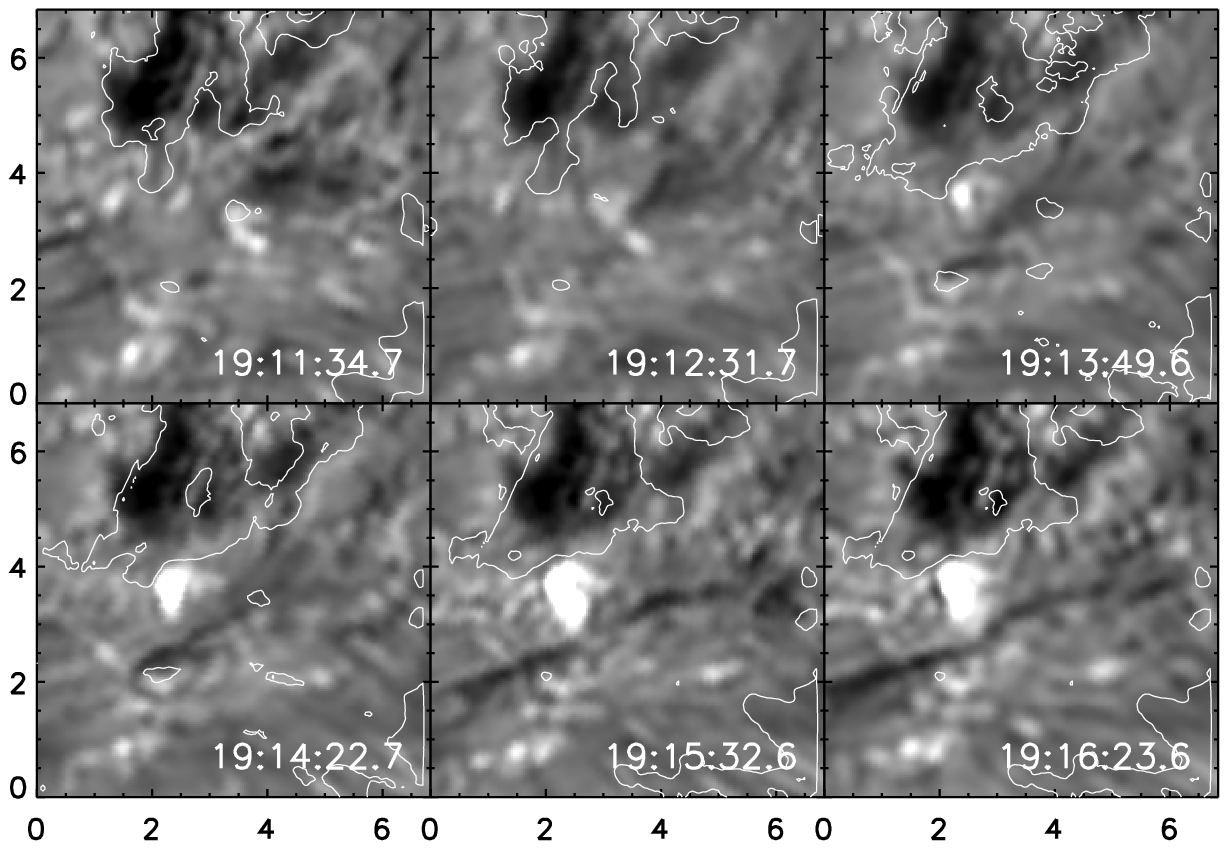}
    \includegraphics[width=0.45\textwidth]{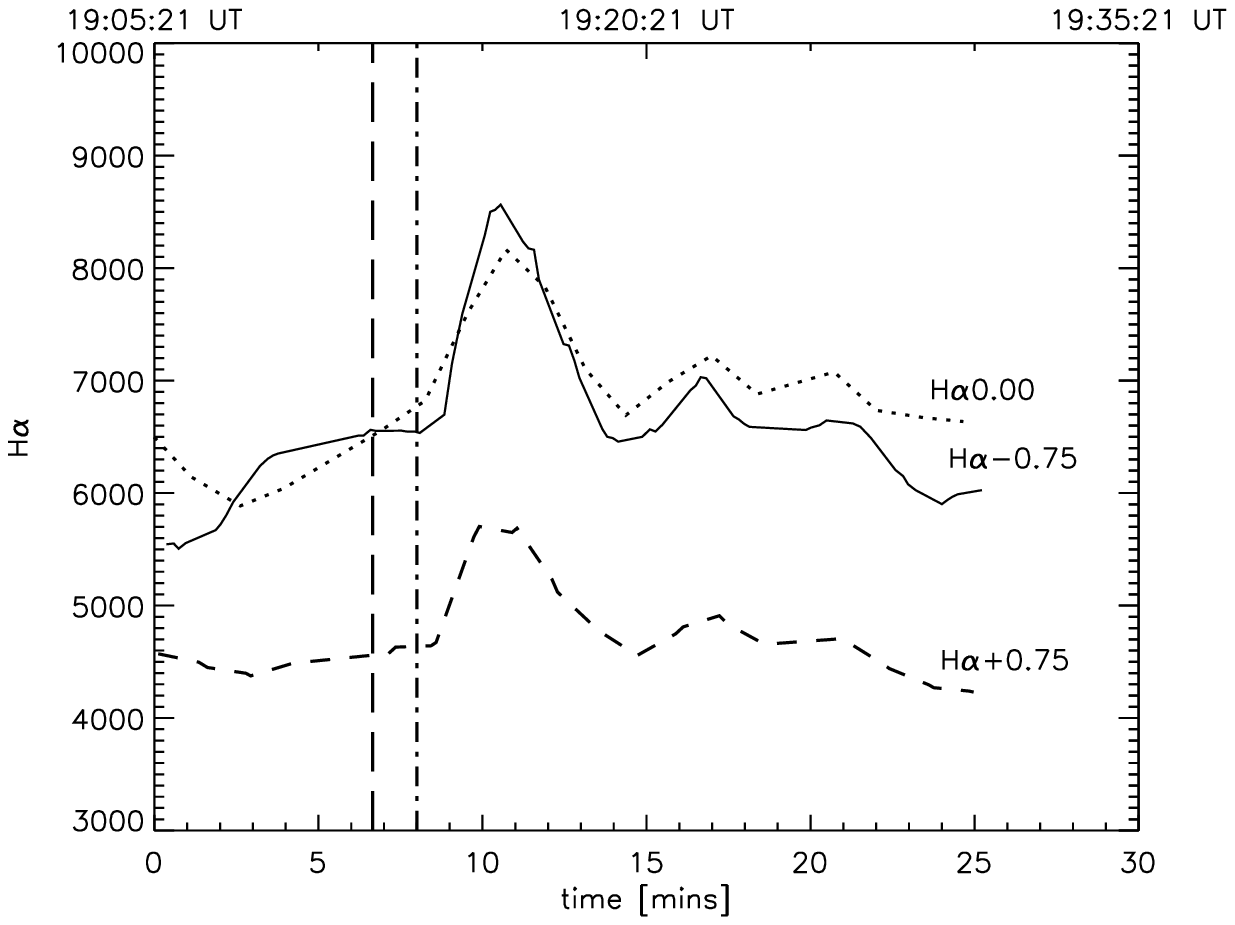}
    \caption{Left: H$\alpha$ intensity maps at $-0.75$~\AA. Right: Time profiles of H$\alpha$ mean intensity measured at the left footpoint of the GLF for all three wavelengths. Vertical long-dash line indicates the time when the positive magnetic flux that is co-spatial with the GLF's left footpoint disappeared, and the vertical dash-dot line indicates the time when H$\alpha$ intensity began to increase.}\label{mmf2_ha.intflux}
\end{center}
\end{figure}

Due to an interruption in the observations, we could not follow photospheric evolution of this GLF. Nevertheless, time coverage of both IRIM and H$\alpha$ data of this event was enough to produce a x-t plot (Figure~\ref{mmf2_timeslice}) using the IRIM data, and a time profile of the H$\alpha$ intensity (Figure~\ref{mmf2_ha.intflux}) for the entire second event. Note that zero-time indicates the same instant in both x-t plots, but the IRIM plot covers a longer period. A comparison of the two plots shows that a trace of the BP at the GLF's moving footpoint (white arrow) is nicely aligned with the positive flux, and the adjacent black trace of the GLF's boundary is located in front of the moving positive flux. The relatively high slope of the trace of the GLF indicates that the GLF was moving at a high speed ranging from 2.5~km~s$^{-1}$ (between $t=0$ to $t=6.75$) to 4.3~km~s$^{-1}$ (between $t=4$ to $t=6.75$). The right panel shows that the positive flux accelerated after it moved away from the main negative penumbra. The observed magnetic element may be considered as the type III MMF with a bit higher velocity. The disappearance of the positive flux at 480~sec could be due to magnetic cancellation through which the field strength became lower than the magnetic sensitivity of IRIM.

We examined time series of H$\alpha$ data to check the chromospheric response to magnetic cancellation. The left panel in Figure~\ref{mmf2_ha.intflux} shows H$\alpha$ blue-wing images overplotted with contours of Stokes-V signals from IRIM data. The time series of H$\alpha$ images shows that the footpoint of the GLF that was shown as a faint brightening in H$\alpha$ first moved toward and encountered the nearby pore (1st row), and later a compact brightening of significant intensity, as compared to the background, appeared just at the boundary of the pore (2nd row) where the GLF was approaching. The temporal relationship between magnetic cancellation and the H$\alpha$ brightening could be more clearly found from the time profile of the H$\alpha$ intensity (right panel). The H$\alpha$ mean intensity obtained at the moving footpoint of the GLF begins to increase at $t=8$ at all wavelengths (dash-dot vertical line), which is just after the positive flux encountered the negative pore and disappeared (long-dash vertical line). This result suggests that the appearance of the H$\alpha$ brightening may have caused by the magnetic cancellation.

Moreover, a dark jet first appeared at around 19:13:49~UT, then developed into a 5\arcsec\, long feature. Although the location of the jet seemed to be removed from the H$\alpha$ brightening at first, the disagreement could be understood if one takes into account the geometry of the field at this location. Both the brightening and upward dark jet are strong indicators of magnetic reconnection in the chromosphere. A comparison between TiO, IRIM, and H$\alpha$ data strongly supports the idea that the magnetic cancellation observed in event 2 is due to the magnetic reconnection between the small-scale magnetic field of the GLF and the larger-scale magnetic field of the pore.

\section{Conclusions and Discussions}
We observed the formation of elongated GLFs just outside a pore and a sunspot in an active region NOAA 11109. The high spatial and temporal resolution data were taken by the NST at BBSO with the aid of an adaptive optics under very good seeing conditions. Magnetic and chromospheric features that are associated with GLFs were examined using NST/IRIM and H$\alpha$ data along with {SDO}/HMI data. Among several events of GLFs that we found in this active region, two events showed similar signatures in TiO images, however they displayed greatly different behavior in the H$\alpha$ line.

The most significant findings are : 1) both elongated GLFs were preceded by elongated darkening and showed a sharp, dark boundary at their moving front. The footpoint behind that boundary was brighter than its surroundings; 2) one of the two events was large enough (3\arcsec) to show a fine structure consisting of thin, dark and bright threads, which to the best of our knowledge were not reported before; 3) the footpoints of GLFs that have a magnetic polarity  opposite to that of the nearby sunspot (pore), and are located farther from the sunspot (pore), moved away from the sunspot with a relatively high speed of 5.0~km~s$^{-1}$ and 4.3~km~s$^{-1}$, respectively.

Our conclusions are summarized as below.
\begin{enumerate}
    \item   We observed the development of small-scale elongated GLFs on sub-granular scale less than 1\arcsec\ in extent at the tip of penumbral filaments. The footpoints of GLFs were co-spatial with newly emergent moving magnetic flux with the footpoint farther from the sunspot having magnetic polarity opposite to that of the sunspot. The observed properties of these photospheric fine-scale features were consistent with emerging $\Omega$-shaped magnetic loops.
    \item   In both observed events, after 1--2 Mm of displacement of the moving footpoints, the magnetic flux at moving footpoints cancelled out pre-existing opposite flux. A similar photospheric evolution of these events showed very different associated chromospheric dynamics. We speculate that the reason may lie in the different topology of the associated fields.
\end{enumerate}

Magnetic elements in the photosphere have mostly been studied using observations obtained with a G-band filter, since  magnetic elements in G-band images are seen as BPs with a contrast much higher than that of the surrounding granules \citep{Ber95}. Many studies have reported that the BPs observed in G-band are indicative of small-scale flux emergence into the photosphere and also frequently associated with chromospheric events, such as brightenings in H$\alpha$ or CaII images \citep{Ber01}. The TiO $705$~nm line used to obtain our photospheric dataset is known to be temperature-sensitive, too, and as a good diagnostic line for imaging of the umbral regions \citep{Ber03}. Recent observations employing a TiO broadband filter on the BBSO/NST reported detection of tiny BPs that might be related to small-scale magnetic fields \citep{Goo10b, Abr10, And10}. The comparison between co-temporal and co-spatial G-band and TiO image revealed that TiO data show the same features as G-band \citep{Abr10}. Given the size of magnetic elements on sub-granular scale, 0\arcsec.5--1\arcsec\, in our case, and their weak magnetic flux, BPs' magnetic field is often not detected with the current state-of-the-art magnetographs, and thus photospheric intensity maps, such as those obtained with G-band or TiO-lines may be more suitable for studying the dynamics of tiny magnetic elements. In our case, the width and length of one of the GLFs was about 0.5\arcsec\, and 1\arcsec\,, respectively. The associated magnetic field was not detected in a {SDO}/HMI magnetogram, but the NST/IRIM data showed the Stokes-V signal even with the loss due to the cross-talk.

Observations of elongated GLFs as a manifestation of flux emergence in an active region were previously reported by \citet{Str99} and \citet{Sch10} who studied an emergence of an active region and an area immediately adjacent to the sunspot penumbra, respectively. \citet{Sch10} measured the Stokes profiles along the elongated granule to show the existence of a horizontal field on the granule's spine and the vertical field of two footpoints, supporting the idea of the emergence of an $\Omega$--shaped loop. Comparing to their works, our data show the entire evolution of a clear morphology of elongated GLFs with the unprecedented spatial and temporal resolution that has not been achieved from other observational data so far. Such observations on the morphology of the photospheric small-scale features could give us an important clue to understanding the interaction between emerging magnetic field and the granular convective flows.

Recent simulations on flux emergence show quite realistic photospheric phenomena, such as the formation of elongated granules and their associated transient darkening or BPs \citep{Che08, Che10}. Our observed data show even more complex structures, such as dark and bright threads, a dark boundary at the moving front, and as well as a relatively high speed of migration of the footpoints of GLFs. Those complex structures of the emerging flux may indicate that even the small-scale flux is already fragmented. This finding was possible due to high spatiotemporal resolution. The time duration of the GLF's development was only $\sim 20$ minutes, and the timespan during which we could see these fine threads was only $\sim 5$ minutes. Therefore, low time-cadence may not catch such a development of fine structures. Also, the mixture of thin, dark and bright threads may be smeared out with the lower spatial resolution.

Observed fine-threads from the first GLF are similar to those of penumbral filaments. Moreover, both of two events occurred at tips of penumbral filaments and were associated with MMFs at relatively high speed compared to that of the moat flow. These findings lead us to fundamental question on the formation mechanism of GLFs and their relation with penumbral magnetic field. Although it is not certain whether the observed GLFs are direct extensions of attached penumbral filaments or not, the possible relation between developing GLFs and forming penumbral filaments is worth considering. In the moving flux tube model suggested by \citet{Sch02}, magnetic field obtains the sea-serpent configuration as outward migrating waves are developed. \citet{Wei04} suggested that the `turbulent pumping by granular convection' drags flux tube downward in the moat region to form sea-serpent configuration. Such a magnetic configuration is frequently adopted to interpret observation properties of MMFs and their relation with penumbral structures \citep{Sai05, Sai08}. Our observed MMFs and GLFs could be also explained by the emergence of migrating $\Omega$-shape crests of sea-serpent field and associated intensity enhancement due to radiative cooling.

Not only the formation of elongated GLFs themselves, but also the observation of two different cancellation phenomena merits discussion. Despite the similar photospheric observational aspects of formation of GLFs, the first one cancelled out without much of a noticeable chromospheric event, while the other one produced a significant chromospheric brightening and a jet. A brightening was also detected in {SDO}/AIA 1600~\AA\, images, while no signatures were observed in other wavelengths. This may indicate the magnetic reconnection probably involved small-scale field lines lying at a chromospheric height. Along with such brightening in both chromospheric H$\alpha$ and EUV lines, the dark jet in H$\alpha$ line is the typical indications of magnetic reconnection. Note that there is a negative pore in the upper-left corner of each panel in Figure~\ref{mmf2_tio.irim.hal}, and the positive polarity of the GLF moves toward this pore. Dark fibrils lining up toward the lower-left from the pore seen in H$\alpha$ centerline images indicate that the magnetic field of the negative pore connects to the positive polarity farther away from the area of interest. The direction of the dark jet that is seen in the H$\alpha$ blue-wing images coincides with the field direction indicated by fibrils. This is consistent with a picture of magnetic reconnection occurring between a pre-existing large-scale field and a newly emerging $\Omega$--shape loop of which one magnetic footpoint approaches the pre-existing opposite polarity. Such reconnection may produce both a dark jet with cool material ejecting upwards following the direction of the large-scale field lines and a brightening in either part of the approaching loops \citep{Yur10}. The converging motion of the two opposing magnetic polarities is often believed to play an important role in driving magnetic reconnection \citep{Chae02,von06}. The movement of the farther magnetic footpoint of observed GLFs was quite fast reaching $4$--$5$~km~s$^{-1}$. That rapid converging motion may have driven magnetic reconnection producing significant brightening in H$\alpha$ in spite of the small size of magnetic flux. Such fast speed is not unexpected, and \citet{Chae10} found a converging speed of H$\alpha$ bright footpoints at 4.6~km~s$^{-1}$ before magnetic reconnection.

The absence of noticeable chromospheric activity associated with the first event suggests that it is possible that the disappearance of two closely located opposite polarity elements may not always be due to magnetic reconnection, but rather due to either emergence of U-loops or submergence of $\Omega$-loops. The first event showed both the appearance of positive magnetic polarity (P1) and its disappearance with the pre-existing negative polarity (N2). Note that the protrusion of negative flux (N1) from the penumbra was also observed when P1 appeared. We carefully suggest that this phenomenon may represent emerging $\Omega$-loops based on the magnetic configurations, such as the nearest footpoint having the same magnetic polarity as the pore. And also, this event was associated with the Doppler blue-shift taken by SDO/HMI with the value of about $-890$~m~s$^{-1}$ when average quiet region was set to zero. Magnetic reconnection could also occur without producing observable jet, and such possibility may not be fully ruled out. Still, it seems possible to suggest an additional scenario that could explain observed features. In case of the disappearance of P1, N2 was comparable to P1 both in size ($\sim2\arcsec$) and in magnetic flux ($\sim 10^{18}$Mx). Such a small flux is likely to be a part of sea-serpent field near the surface forming the sequence of U- and $\Omega$-shaped loops. Moreover, the H$\alpha$ intensity at P1 showed its maximum before P1 encountered with N2. Based on that idea, we suggest an additional explanation for this event, such as the magnetic field of the newly emerging bipole and that of the pre-existing negative pole may be connected below the surface in the form of the sea-serpent field, and the emergence of such a field could result in the emerging bipole and the subsequent cancellation.

\acknowledgments We are grateful to an anonymous referee for the valuable comments and suggestions that significantly improved the paper. We also appreciate J.Chae for a careful reading of this manuscript and comments, and A.Sainz Dalda for the helpful comments. Authors thank BBSO observers and the instrument team for their contribution to this study. E.-K.L. and P.G. are partially supported by AFOSR (FA9550-09-1-0655). V.Y.'s work was partly supported under NASA GI NNX08AJ20G and LWS TR\&T NNG0-5GN34G grants. V.A. acknowledges partial support from NSF grant ATM-0716512. P.G., V.A. and V.Y are partially supported by NSF (AGS-0745744) and NASA (NNY 08BA22G). K.A. and W.C. acknowledge support from NSF Grant AGS-0847126.

\end{document}